\documentclass[10pt]{article}

\usepackage[svgnames]{xcolor}
\usepackage[colorlinks,citecolor=DarkGreen,linkcolor=DarkRed,urlcolor=DarkBlue]{hyperref}
\usepackage{amsmath,amssymb,dsfont,slashed,color,hyperref}
\usepackage{mathtools}
\usepackage{authblk}
\usepackage[toc,page]{appendix}
\usepackage{makecell}

\definecolor{armygreen}{rgb}{0.29, 0.33, 0.13}

\newcommand{\half}{\frac{1}{2}}

\newcommand{\JJ}{\mathds{J}}
\newcommand{\KK}{\mathds{K}}
\newcommand{\DD}{\mathds{D}}
\newcommand{\TT}{\mathds{T}}

\newcommand{\ZZ}{\mathds{Z}}

\newcommand{\QQ}{\mathds{Q}}
\newcommand{\QQb}{\overline{\mathds{Q}}}

\newcommand{\AAA}{\mathds{A}}
\newcommand{\FF}{\mathds{F}}
\newcommand{\calF}{\mathcal{F}}
\newcommand{\calG}{\mathcal{G}}
\newcommand{\calH}{\mathcal{H}}

\newcommand{\se}{\slashed{e}}

\newcommand{\sD}{\slashed{D}}

\newcommand{\lsD}{\overleftarrow{\sD}}
\newcommand{\psibar}{{\overline{\psi}}}

\newcommand{\gf}{\gamma_5}
\newcommand{\dV}{|e|d^4x}

\newcommand{\MP}{{M_\text{P}}}
\newcommand{\GN}{{G_\text{N}}}

\newcommand{\calX}{\mathcal{X}}
\newcommand{\ocalX}{\overline{\mathcal{X}}}

\newcommand{\cecs}{Centro de Estudios Cient\'{\i}ficos (CECs), Arturo Prat 514, Valdivia, Chile.}

\newcommand{\ubb}{Centro de Ciencias Exactas \& Departamento de Ciencias B\'asicas, Universidad del Bio-Bio, Avda. Andrés Bello 720, Casilla 447 - CP: 3800708, Chill\'an, Chile.}

\newcommand{\ussvaldivia}{Facultad de Ingenier\'ia, Universidad San Sebasti\'an, General Lagos 1163, Valdivia, Chile}

\begin{document}

\title{Chiral symmetry breaking in models with unconventional supersymmetry}
\author[1,3]{P. D. Alvarez \thanks{E-mail: \href{mailto:pedro.alvarez@uss.cl}{\nolinkurl{pedro.alvarez@uss.cl}}}}

\author[2]{Cristian Villavicencio \thanks{E-mail: \href{mailto:xxx@yyy.zzz}{\nolinkurl{cvillavicencio@ubiobio.cl}}}}

\author[1,3]{J. Zanelli \thanks{E-mail: \href{mailto:z@cecs.cl}{\nolinkurl{z@cecs.cl}}}}

\affil[1]{\ussvaldivia}
\affil[2]{\ubb}
\affil[3]{\cecs}

\maketitle

\begin{abstract}
We investigate dynamical mass generation in a geometrically constructed gauge theory based on the super Lie algebra $su(2,2|3)$, in which gravity, Yang-Mills fields, and fermions are unified as components of a single gauge connection. The model contains no elementary scalar fields and no ad hoc four-fermion interactions. Instead, fermionic self-interactions arise unavoidably from the geometric structure of the theory, through nonminimal couplings and torsion associated with the unified connection. Upon reduction to an effective low-energy description, these interactions generate a Nambu-Jona-Lasinio--type potential that triggers chiral symmetry breaking and the formation of a fermion mass gap. In this framework, mass generation emerges as a direct consequence of the underlying gauge-geometric and algebraic structure, rather than as an independent dynamical assumption.
\end{abstract}


\section{Introduction}\label{sec:intro}

One of the central open problems in fundamental physics is the origin of mass. In the Standard Model, particle masses arise through spontaneous symmetry breaking driven by an elementary scalar field  \cite{Higgs:1964pj,Englert:1964et}, while in strongly coupled theories such as QCD, masses can be generated dynamically through fermion condensation \cite{Nambu:1961tp,Hatsuda:1994pi}. These mechanisms are conceptually distinct, yet both rely on introducing specific dynamical ingredients whose deeper origin is not determined within the theory.

From a complementary perspective, geometric and gauge-theoretic approaches to gravity \cite{Kibble:1961ba,Hehl:1976kj} suggest that spacetime geometry and internal symmetries may be unified within a single mathematical structure \cite{MacDowell:1977jt,VanNieuwenhuizen:1981ae}. In such formulations, gravity is treated on the same footing as gauge interactions, often as part of an enlarged gauge symmetry. This raises a natural question: can mass generation itself emerge from geometry and gauge symmetry, without introducing additional scalar fields or ad hoc interactions?

In this work, we address this question in a framework where gravity, Yang-Mills fields, and fermions are all components of a single gauge connection. The underlying symmetry is governed by a super Lie algebra that unifies spacetime and internal degrees of freedom \cite{Alvarez:2021zsw}. In this setting, the fermionic sector contains no explicit mass terms, and the action is fully determined by gauge invariance and geometric consistency.

A key observation is that, once all fields are embedded into a unified connection, fermionic self-interactions arise unavoidably. These interactions are not introduced by hand but follow directly from the geometric couplings of the theory, including nonminimal terms and torsion associated with the gauge. 
When the theory is reduced to an effective low-energy description, these interactions take the form of a four-fermion potential connection \cite{Castillo-Felisola:2013jva} closely related to
the Nambu--Jona-Lasinio mechanism \cite{Nambu:1961tp}.

This effective interaction triggers chiral symmetry breaking and leads to the dynamical generation of a fermion mass gap. In this sense, mass generation emerges as a consequence of the gauge-geometric structure of the theory rather than as an independent dynamical assumption. The resulting mechanism shares qualitative features with dynamical mass generation in QCD, while being rooted in a fundamentally different, geometric origin.

The specific model studied here is based on the super Lie algebra $su(2,2|3)$, which naturally incorporates four-dimensional gravity together with an internal $SU(3)$ gauge symmetry and fermionic matter. This algebraic structure fixes the relative couplings of the theory and strongly constrains the allowed interactions. As a result, the appearance of the effective four-fermion terms and the ensuing mass generation mechanism are highly rigid features of the construction.

While the fermion implied above may naturally be interpreted as a QCD-like quark in a strongly interacting sector, the same framework can also be viewed from a broader perspective. In particular, the fermionic degree of freedom need not belong to the visible sector and may instead constitute part of a hidden sector with potential dark matter applications. This alternative interpretation is especially appealing because the underlying geometric construction constrains the interactions of the fermion in a nontrivial way, providing a predictive setting for exploring dark-sector phenomenology.

This rigidity is important from a phenomenological point of view. Since minimal single-component dark matter scenarios are now strongly constrained, it is natural to consider more elaborate dark sectors. However, enlarging the field content and the number of couplings can easily reduce predictivity unless the additional structure imposes relations among the parameters and observables. In this respect, our construction follows a bottom-up logic similar in spirit to simplified dark matter models, which provide a small set of physically transparent parameters while retaining contact with possible ultraviolet completions and allowing complementary constraints from different search channels \cite{DeSimone:2016fbz}. The difference is that, in the present framework, the correlations among couplings are not imposed phenomenologically: they are dictated by the underlying superalgebraic gauge structure. Thus, the model is non-minimal, but not arbitrary. Its effective dark-sector parameters, including the dynamically generated fermion mass and the strength of the induced four-fermion interaction, arise from a common geometric origin and are therefore tied to one another.

This paper is organized as follows. In Sec.~\ref{sec:geometricmodel} we present the geometric model and in Sec.~\ref{sec:model} its reduction to an effective low-energy theory. In Sec.~\ref{sec:massgap} we analyze the resulting four-fermion interactions and derive the mass gap equation. 
In Sec.~\label{sec:compositemass} we explore the properties of composite states and phenomenological implications. We conclude in Sec.~\ref{sec:conclu} with a discussion of the broader significance of geometric mass generation and possible extensions.

\section{Gauge-Geometric Construction}\label{sec:geometricmodel}

Inspired by the MacDowell--Mansouri formulation of supergravity \cite{MacDowell:1977jt}, we proposed in \cite{Alvarez:2021zsw} a geometrically constructed gauge model based on the action
\begin{equation}\label{action}
 \mathcal{S}= - \int \langle \FF \circledast \FF \rangle\,,
\end{equation}
The theory is formulated in terms of a superalgebra--valued gauge connection whose curvature encodes both gravitational and internal degrees of freedom.

The gauge connection takes the form
\begin{equation}\label{gaugeconnection}
\AAA= \Omega+\QQb_\alpha^i \se \psi_i^\alpha+\overline{\psi}_\alpha^i \se \QQ_i^\alpha\ \,,
\end{equation}
where $\psi_i^\alpha$ are fermionic zero-forms transforming in the fundamental representation of $SU(N)\times U(1)$, and $\overline{\psi}_\alpha^i$ denotes their Dirac conjugate. The bosonic sector of the connection is contained in
\begin{equation}
\Omega= \half \omega^{ab}\JJ_{ab}+f^a \JJ_a+g^a\KK_a+h\DD+A^I \TT_I+A\ZZ \,.
\end{equation}
Here $\omega^{ab}$ is the Lorentz connection, $f^a$ and $g^a$ are vector-valued one-forms associated with the translational and conformal generators, $h$ corresponds to dilatations, and $A^I$ and $A$ are the $SU(N)$ and $U(1)$ gauge fields, respectively.

The associated field strength is defined in the usual way as
\begin{equation}
 \FF = d\AAA +\AAA \wedge \AAA\,,
\end{equation}
To construct the action \eqref{action}, we introduce a generalized dual operator $\circledast$ that partially breaks the $SU(2,2|N)$ symmetry. While a more general class of such operators was discussed in \cite{Alvarez:2021zsw}, for the purposes of the present work we consider
\begin{align}
\circledast \FF=&S \left(\half \calF^{ab} \JJ_{ab}+\calF^a \JJ_a +\calG^a \KK_a\right)\nonumber\\
&+\ast\calH \DD +\ast \calF^I \TT_I +\ast\calF \ZZ + \QQb (-i\gf) \calX + \ocalX (-i\gf ) \QQ\,.\label{dualop}
\end{align}
where $\ast$ denotes the spacetime Hodge dual and $\gf$ is the chirality matrix. Explicit expressions for the curvature components appearing above are collected in Appendix~\ref{AppA}.

The resulting model possesses $SO(1,3)\times SU(N)\times U(1)\subset SU(2,2|N)$ gauge symmetry. An important feature of the construction is that the fields $f^a$ and $g^a$ enter the action without kinetic terms and therefore play the role of auxiliary geometric fields. Consistent solutions of the field equations are obtained by fixing the background geometry according to
\begin{equation}\label{auxiliaryfields}
 f^a = \rho e^a\,, \qquad g^a = \sigma e^a\,, \quad \rho > \sigma\,.
\end{equation}
where $e^a$ is the vierbein.

This choice is compatible with the equations of motion and reduces the geometric sector to an anti--de Sitter background with nonvanishing torsion. Upon this reduction, the model effectively yields a $SU(N)\times U(1)$ Yang-Mills theory coupled to gravity and fermionic matter, with the relative couplings fixed by the underlying geometric structure.

\section{Phenomenological Model}\label{sec:model}

We now consider the phenomenological model obtained by restricting the geometric construction of Ref.~\cite{Alvarez:2021zsw} to the case of $N=3$ colors and reducing the theory to the anti--de Sitter sector in a generic non-chiral configuration. The resulting effective action takes the form
\begin{align}\label{action}
 \cal{S} &=&  \int \dV \left[\frac{1}{16\pi \GN}(R-2\Lambda)-\frac{1}{4}H^{\mu\nu}H_{\mu\nu}-\frac{1}{4}F^{I\mu\nu}F^{I}_{\mu\nu}-\frac{1}{4}F^{\mu\nu}F_{\mu\nu} \right.   \nonumber\\
   && \left. +\frac{1}{2}
  \psibar (\lsD-\sD)\psi  + T_\text{NMC}\right]
\end{align}
where $T_{\text{NMC}}$ denotes nonminimal coupling terms and $D$ is the covariant derivative associated with the $SO(3,1)\times SU(3)\times U(1)$ gauge symmetry.

The explicit form of the covariant derivatives acting on spinors is
\begin{align}
 (D)^{\alpha j}_{i \beta}&=\delta^j_i \delta^\alpha_\beta d +\half \omega^{ab}\delta^j_i(\Sigma_{ab})^\alpha_{\ \beta}-\frac{i}{2}g^{(SU(3))}A^I(\lambda_I)_i^{\ j}\delta^\alpha_\beta-i g^{(U(1))}A\delta^j_i \delta^\alpha_\beta\,,\label{covD}\\
 (\overleftarrow{D})^{\alpha j}_{i \beta}&=\overleftarrow{d}\delta^j_i \delta^\alpha_\beta -\half \omega^{ab}\delta^j_i(\Sigma_{ab})^\alpha_{\ \beta}+\frac{i}{2}g^{(SU(3))}A^I(\lambda_I)_i^{\ j}\delta^\alpha_\beta+ig^{(U(1))}A\delta^j_i \delta^\alpha_\beta\,.\label{covDl}
\end{align}

The underlying superconformal symmetry fixes the relative values of all coupling constants appearing in the action. In particular,
\begin{align}
\Lambda = & -\frac{3M_p^2}{2\xi}\,,\\
g^{(SU(3))} = &\frac{1}{\sqrt{\xi}}\,,\\
g^{(U(1))} = &\frac{1}{2\sqrt{6\xi}}\,.
\end{align}

In Eq.~\eqref{action} we have suppressed spinor and $SU(3)$ fundamental indices for notational simplicity. For example,
\begin{equation}
 \psibar \psi = \psibar^i_\alpha \psi_i^\alpha\,, \qquad \psibar \slashed{D}\psi = \psibar^i_\alpha (\gamma^\mu D_\mu)^{\alpha j}_{i\beta}\psi_j^\beta \qquad \psibar \overleftarrow{\slashed{D}}\psi = \psibar^i_\alpha ( \overleftarrow{D}_\mu)^{\alpha j}_{i\beta}\gamma^\mu\psi_j^\beta\,.
\end{equation}
Our index conventions are summarized in Table~\ref{indices}.

\begin{center}
\begin{table}[]
    \centering
\begin{tabular}{lll}
description & index & range\\ \hline
curved spacetime & $\mu, \nu $& $0, 1, \cdots, 3$\\
tangent space & $a, b, \cdots$ & $0, 1, \cdots, 3$\\
$SU(3)$ adjoint & $I,J,\cdots $& $1,2,\cdots,8$\\
$SU(3)$ fundamental & $i,j$ & $1,2,3$\\
Spinor  & $\alpha,\beta$ & $1,\cdots,4$
\end{tabular}
\caption{Summary of our index conventions.}\label{indices}
\end{table}
\end{center}

The nonminimal coupling term $T_{\text{NMC}}$ includes couplings to torsion, a gravitational Pauli-like interaction, and quartic fermion terms,
\begin{align}
 T_\text{NMC}=&\frac{i}{6} T_{bcd}\epsilon^{abcd} \psibar \gf \gamma_a \psi - \frac{\sqrt{2\xi}}{3\MP}R^{ab}{}_{ab}\psibar \psi\nonumber\\
 &+\frac{1}{12 M_P^2}\left(\phi^2+\phi^2_5-\frac{1}{18}\phi^{ab}\phi_{ab}+\frac{1}{6} \phi^{abI}\phi_{abI}\right)
 \,,\label{NMC}
\end{align}
where $R^{ab}{}_{cd}$ is defined by
\begin{equation}
    R^{ab} = \frac{1}{2}R^{ab}{}_{cd} e^c e^d\,,
\end{equation}
and
\begin{equation*}
  \phi=\psibar\psi, \qquad \phi_{ab}=\psibar\gamma_{ab}\psi, \qquad \phi_{5}=\psibar\gamma_{5}\psi,\qquad \phi_{abI}=\psibar\gamma_{ab}\lambda_{I}\psi.
\end{equation*}
The quartic fermion terms are highly suppressed and do not receive contributions proportional to the phenomenological parameter $\xi$. The first term in Eq.~\eqref{NMC} couples the totally antisymmetric component of the torsion two-form, $T^a=\half T^a{}_{bc}e^b e^c$, to the 
 axial fermion bilinear. In the presence of a fermion condensate, this coupling allows the torsion background to be interpreted as an effective chiral chemical potential.

Apart from the non-minimal couplings, the action \eqref{action} corresponds to a standard Einstein-Dirac-Yang-Mills theory with gauge group $SU(3)$, with all relative couplings fixed by the underlying geometric construction. In a constant-curvature background,
\begin{eqnarray}
    R^{ab} - \frac{\Lambda}{3}e^a e^b = 0\,,
\end{eqnarray}
and then the second term in Eq.~\eqref{NMC} induces an effective fermion mass
\begin{eqnarray}
    m_\text{eff} = &\frac{4\sqrt{2} M_p}{\sqrt{\xi}}\,.
\end{eqnarray}

 In the case of the third term in Eq.\,\eqref{NMC}, using the identity $(\lambda_I)_{ij}(\lambda_I)_{kl}= 2(\delta_{il}\delta_{kj}-\delta_{ij}\delta_{kl}/N)$ and applying Fierz rearrangements, the quartic fermion terms reduces to
\begin{equation}
    \frac{1}{6M_P^2}\left[(\bar \psi \psi)^2 + (\bar \psi \gamma_5\psi)^2
    +\frac{1}{18} \left(1-\frac{3}{N}\right)(\bar \psi \gamma_{ab}\psi)^2
    \right].
\end{equation}
For $N=3$, the tensor channel vanishes identically.

Variation with respect to the Lorentz connection gives an algebraic equation expressing torsion in terms of fermion bilinears.
If one subtitutes the last back into the action, additional contributions to the effective Nambu--Jona-Lasinio interaction will appear but in the axial channel.

The Dirac equation therefore contains, in addition to the minimally coupled Dirac operator, a quartic fermion term responsible for chiral symmetry breaking and the dynamical generation of a mass gap. The internal gauge fields $A$ and $A^I$ satisfy the standard Yang--Mills equations in curved spacetime, sourced by the corresponding matter currents.

In the next section we analyze the mass gap generated by the fermion quartic interactions.

\section{NJL analysis}\label{sec:massgap}

The four-fermion interaction emerging from the geometric construction resembles the Nambu--Jona-Lasinio (NJL) model originally introduced in the context of QCD \cite{Nambu:1961tp}. In the present case, the effective theory describes single-flavor fermion dynamics with an $SU(3)$ color symmetry. Accordingly, we consider the four-fermion interaction\footnote{In QCD, the gamma-matrix conventions and fermion fields differ from those used here, namely $\gamma\to i\gamma$ and $\bar\psi\to -\bar\psi$.}
\begin{equation}
    {\cal L}_4= G\left[(\bar \psi \psi)^2 + (\bar \psi \gamma_5\psi)^2\right]
\end{equation}
where the NJL coupling is fixed by the underlying geometric model, $G^{-1}=6M_P^2$. Throughout this section we work in an AdS vacuum with vanishing torsion, $T^a=0$.

The NJL Lagrangian is expected to be invariant under $SU(2)_f$ chiral transformations. In the present setup, however, the effective model should be regarded as a toy model: it contains a single fermion flavor and is invariant under $U(1)$ chiral symmetry. 
Despite these limitations, our goal is to investigate whether this emergent NJL-like interaction can reproduce qualitative low-energy features familiar from QCD, starting solely from fermionic degrees of freedom. In this sense, we are interested in exploring possible low-energy predictions of the
model.

Assuming the validity of the mean-field approximation and the formation of a fermion condensate, the mass gap equation follows from the dressed single-flavor fermion propagator,
\begin{equation}
    M = m_\text{eff} +\frac{1}{M_P^2}\int \frac{d^4p}{(2\pi)^4}\frac{4i M}{p^2-M^2+i\varepsilon} 
    \label{eq:gap}
\end{equation}
where $m_{\text{eff}}$ denotes the bare fermion mass, which in the present model is an effective mass induced by the background geometry, and $M$ is the dynamically generated fermion mass dressed by the condensate. Following standard QCD terminology, we refer to $M$ as the \emph{constituent mass}.

The integral in Eq.~\eqref{eq:gap} is ultraviolet divergent and must be regularized by introducing a cutoff $\Lambda_c$,
\begin{equation}
    \int \frac{d^4p}{(2\pi)^4}\frac{4iM}{p^2-M^2+i\varepsilon} = \frac{1}{\pi^2} \int_0^{\Lambda_c}dp \frac{M p^2}{\sqrt{p^2+M^2}}.
\end{equation}

The model contains two free parameters: the dimensionless parameter $\xi$, which controls the effective mass $m_{\text{eff}}$, and the NJL cutoff $\Lambda_c$. Since our aim is to mimic qualitative features of QCD, the effective mass $m_{\text{eff}}$ should be small. Large values of $\xi$ correspond to small $m_{\text{eff}}$, as well as to suppressed gauge couplings and a small cosmological constant. In the limit $\xi\to\infty$, the theory approaches the chiral limit with decoupled gauge fields, becoming increasingly similar to the standard NJL model.

If the goal is to reproduce meson-like states with realistic masses, the dominant contribution must arise from the NJL interaction and the fermion condensate. 
 In the NJL model, the coupling constant is commonly expressed in terms of the cutoff scale, with naturalness arguments and phenomenological fits indicating that the dimensionless combination 
$G \Lambda_c^2$ is of order unity. 
Since the NJL coupling in the present model is extremely small, $G\sim M_P^{-2}$, the cutoff $\Lambda_c$ must be of the order of the Planck mass. In this regime, the gap equation yields constituent fermion masses of order GeV.

\begin{figure}[h]
  \centering
  \includegraphics[trim={0 0cm 0 0},clip,width=.7\linewidth]{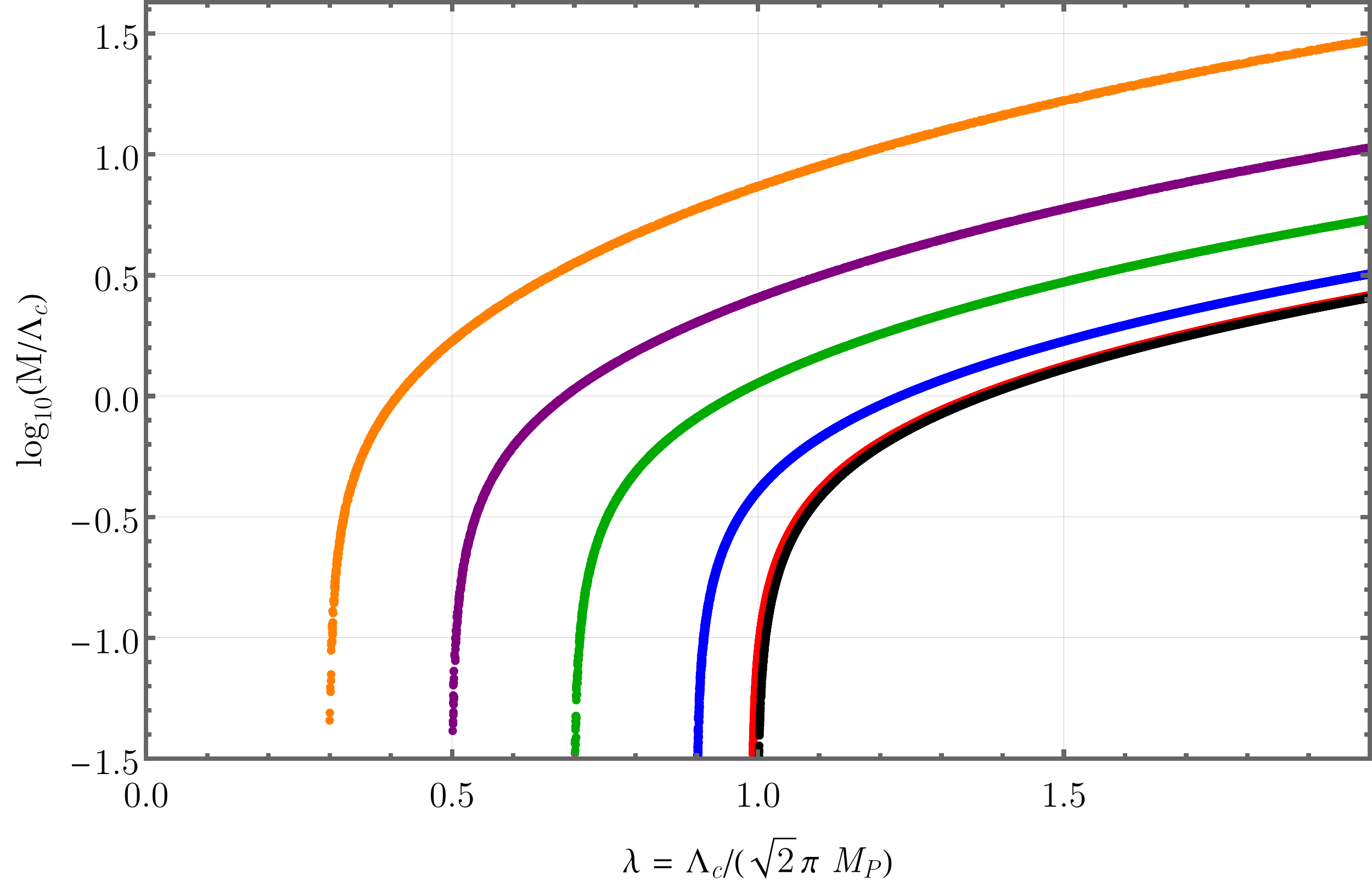}
  \includegraphics[trim={0 0cm 0 0},clip,width=.1\linewidth]{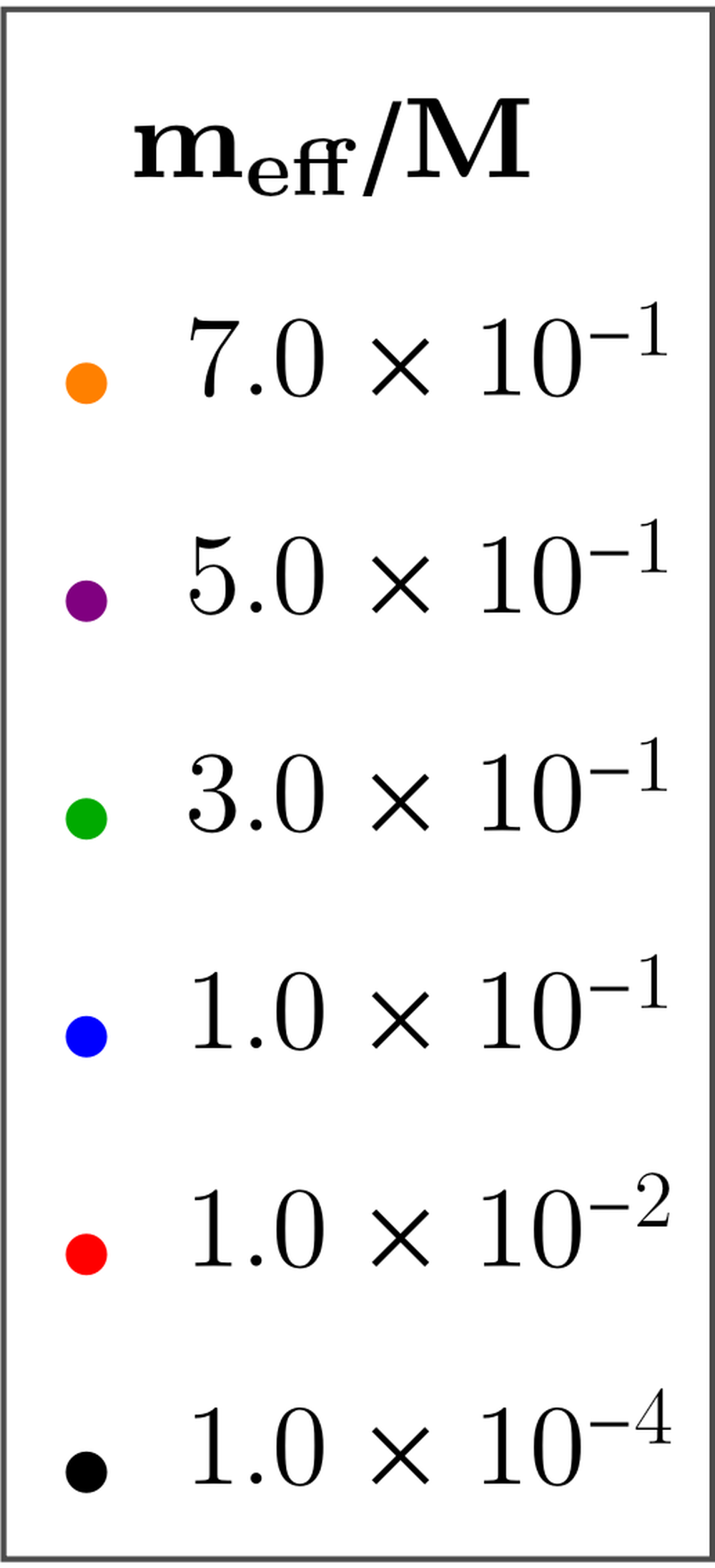}
\caption{Constituent mass as a function of the NJL cutoff for several values of $m_{\mathrm{eff}}$. The case $m_{\mathrm{eff}} \sim 10^{-4}$ is indistinguishable, within the line width, from the chiral limit $m_{\mathrm{eff}} \to 0$.}
\label{mass_vs_cutoff}
\end{figure}

A direct numerical solution of the gap equation in the chiral limit is shown in Fig.~\ref{mass_vs_cutoff}, where the constituent mass is plotted as a function of the cutoff, scaled by the Planck mass. The figure shows that the constituent mass remains zero up to a critical value $\Lambda_c\simeq \sqrt{2}\,\pi M_P$.

Further insight is obtained by introducing the dimensionless parameters
\begin{equation}
    \lambda \equiv \frac{\Lambda_c}{\sqrt{2}\pi M_P},
    \quad   \bar M\equiv \frac{M}{\Lambda_c},
    \quad \bar m_\text{eff} \equiv \frac{m_\text{eff}}{M},
    \label{eq:scaled_param}
\end{equation}
in terms of which the gap equation \eqref{eq:gap} becomes
\begin{equation}
    1 = \bar m_\text{eff} + \lambda^2\left\{\sqrt{1+\bar M^2}+\bar M^2\ln\left[\frac{\bar M}{\sqrt{1+\bar M^2}+1}\right] \right\}.
\end{equation}
Since $\Lambda_c \sim M_P$ while $M$ is expected to be of the order of $1\,\mathrm{GeV}$, the parameter $\bar M$ is negligibly small. 
In this regime, the gap equation simplifies to $\lambda^2 \simeq 1 - \bar m_{\mathrm{eff}}$. 
Consequently, these $\lambda$ values correspond approximately to the intersections with the horizontal axis in Fig.~\ref{mass_vs_cutoff}.

For a typical light-quark constituent mass $M\simeq 340$\,MeV and a current mass $m_{\text{eff}}\simeq 5$\,MeV, one finds $\lambda\simeq 0.985$. While these values are indicative, the determination of meson masses in this framework turns out to be an extremely fine-tuned problem.

For $M\ll M_P$, a useful approximation for the cutoff is
\begin{equation}
    \lambda^2 = 1-\frac{m_\text{eff}}{M}+\frac{M^2}{4\pi^2 M_P^2}\left[
    \ln\left(8\pi^2 M_P^2/M^2\right)-1
    \right].
\end{equation}

As illustrated in Fig.~\ref{mass_vs_cutoff}, achieving constituent masses of order $M\sim 1$\,GeV requires an extraordinarily precise tuning of the cutoff. Quantitatively, this corresponds to $|\lambda-1|\ll 10^{-22}$. In practice, the only well-behaved choice is $\lambda=1$, in which case setting $M=340$\,MeV leads to an effective bare mass $m_{\text{eff}}\sim 10^{-42}$\,MeV, effectively realizing the chiral limit.

\section{Masses of composite particles}\label{sec:masscomposite}

Within the NJL framework, meson masses can be obtained by solving the Bethe--Salpeter (BS) equation in the random phase approximation (RPA) \cite{Schwarz:1999dj,Buballa:2003qv}. In this approach, mesons appear as poles of the four-fermion interaction $T$-matrix in the corresponding channel,
\begin{equation}
    T_{\cal M}(p^2)=\frac{2G}{1-2G\Pi_{\cal M}(p^2)} \sim \frac{1}{p^2-m_{\cal M}^2}
\end{equation}
where $\Pi_{\mathcal{M}}(p^2)$ is the polarization function in the meson channel $\mathcal{M}$, and the pole position defines the meson mass $m_{\mathcal{M}}$.

The polarization function is given by
\begin{equation}
    \Pi_{\cal M}(p^2) = -i\int\frac{d^4k}{(2\pi)^2}\text{ tr}\left[{\cal O}_{\cal M} S(p+q){\cal O}_{\cal M}S(k)\right]
\end{equation}
where $S(k)$ denotes the dressed fermion propagator in the mean-field approximation and $\mathcal{O}_{\mathcal{M}}$ is the operator associated with the mesonic channel under consideration. In this work we focus on the pseudoscalar and scalar channels, corresponding to $\mathcal{O}_p=i\gamma_5$ and $\mathcal{O}_s=\mathds{1}$, respectively.

The introduction of a momentum cutoff explicitly breaks Lorentz invariance, so the polarization function depends separately on energy and spatial momentum, $\Pi_{\mathcal{M}}=\Pi_{\mathcal{M}}(p_0^2,\boldsymbol{p}^2)$. Consequently, the BS equation must be solved in the meson rest frame in order to extract the pole masses,
\begin{equation}
1 = 2G\Pi_{\cal M}(m_{\cal M}^2,0).    
\end{equation}
Using the scaled variables introduced in Eq.~\eqref{eq:scaled_param}, the BS equation for the pseudoscalar channel takes the form
\begin{equation}
    1=2\lambda^2\int_0^1 dk\frac{k^2\sqrt{k^2+\bar M^2}}{k^2+\bar M^2(1-\bar m_p^2)}\,,
\end{equation}
while for the scalar channel one finds
\begin{equation}
    1=2\lambda^2\int_0^1 \frac{dk k^4}{\sqrt{k^2+\bar M^2}}\frac{1}{k^2+\bar M^2(1-\bar m_s^2)}\,.
\end{equation}
Here we have defined the dimensionless meson mass
\begin{equation}
    \bar m_{\cal M} \equiv \frac{m_{\cal M}}{2M}.
\end{equation}
This definition is convenient because the BS equation admits  solutions only for $m_{\mathcal{M}}<2M$. When this condition is violated, the solution becomes complex, signaling an unstable resonance whose imaginary part corresponds to the decay width. In the following, we therefore restrict our analysis to the range $0<\bar m_{\mathcal{M}}<1$.

Although both integrals can be evaluated analytically, the explicit expressions are not particularly illuminating and will not be displayed here. As discussed in the previous section, the analysis is delicate because $\bar M\sim 10^{-22}$, requiring an expansion in $\bar M$ after performing the integrations. Moreover, physically meaningful solutions to the BS equation are obtained only for $\lambda\simeq 1$, and in the numerical analysis we therefore set $\lambda=1$.

For a constituent fermion mass $M=340$\,MeV, we find a pseudoscalar bound-state mass
\begin{equation}
    m_p = 503.2~\text{MeV},
\end{equation}
while no real solution is found in the scalar channel.

\subsection*{Gluequark dark matter paradigm}

In the last example, the model allows, through dynamical symmetry breaking, the formation of a stable pseudoscalar particle within the typical mass range of the Standard Model. 
Although this is a toy model constructed with one-flavor quarks, the results are consistent with the emergence of particles both within and beyond the Standard Model range.

In other contexts, such as dark matter scenarios, this construction may also provide interesting insights into possible dark matter candidates.

Conventional light dark matter scenarios face increasingly severe tensions when embedded in a thermal freeze-out framework. For sub-GeV masses, achieving the observed relic abundance typically requires sizable annihilation rates at late times, which are strongly constrained by cosmic microwave background limits on energy injection from dark matter annihilation, as well as by beam-dump experiments and flavor observables~\cite{Planck:2018vyg}. As a result, viable models of light thermal dark matter often rely on highly suppressed portals, secluded sectors, or finely tuned annihilation channels, significantly complicating their theoretical structure. These challenges motivate alternative frameworks in which the relic abundance is fixed at high temperatures while subsequent low-energy interactions are dynamically suppressed. Gluequark dark matter provides such a mechanism: freeze-out occurs at the level of heavy, weakly coupled constituents prior to confinement, while the later formation of neutral composite states naturally evades low-energy and cosmological constraints without requiring ad hoc symmetries or tuned interactions.

Let us summarize the main features of gluequark dark matter (GQDM) models~\cite{Mitridate:2017izz,Contino:2018crt}:
\begin{enumerate}
    \item Thermal freeze-out occurs while the heavy colored fermions $Q$ are     still free particles at high temperatures, $T \gg \Lambda_D$, where $\Lambda_D$ denotes the confinement scale of the dark gauge group.

    \item As the temperature drops below $T \sim \Lambda_D \ll m_Q$, the dark sector undergoes confinement and the free fermions $Q$ bind into composite states, known as gluequarks.

    \item The comoving number density of the heavy fermions $Q$, fixed at freeze-out, is transferred to the number density of gluequarks, assuming that no further significant annihilation processes occur during the confinement transition.
\end{enumerate}

In the gluequark dark matter paradigm, the dark matter candidate is a stable composite bound state formed by a heavy colored fermion and a dark gluon. This scenario can be naturally embedded within theories of \emph{unconventional supersymmetry}, in which matter fields transforming effectively in fundamental representations emerge from adjoint representations of a larger symmetry algebra. Such constructions extend the standard supersymmetric framework while preserving its essential structural features, and allow heavy colored fermions $Q$ to arise as fermionic components of gauge multiplets rather than from conventional chiral superfields. Their stability may then follow from accidental or residual symmetries, without the need to impose an \emph{ad hoc} $Z_2$ parity.

The presence of a new confining non-Abelian gauge interaction in the dark sector naturally leads to the formation of composite states such as gluequarks. In unconventional supersymmetric settings, gauge fields and fermions can be unified within the same supermultiplets, while the scalar sector is not subject to the usual mass-degeneracy constraints. This structure allows for a flexible hierarchy between the heavy fermion mass $m_Q$, the confinement scale $\Lambda_D$, and the masses of composite and excited states, and provides a consistent framework for describing the non-perturbative dynamics associated with confinement and bound-state formation.

\begin{figure}[h]
  \centering
  \includegraphics[trim={0 0cm 0 0},clip,width=.7\linewidth]{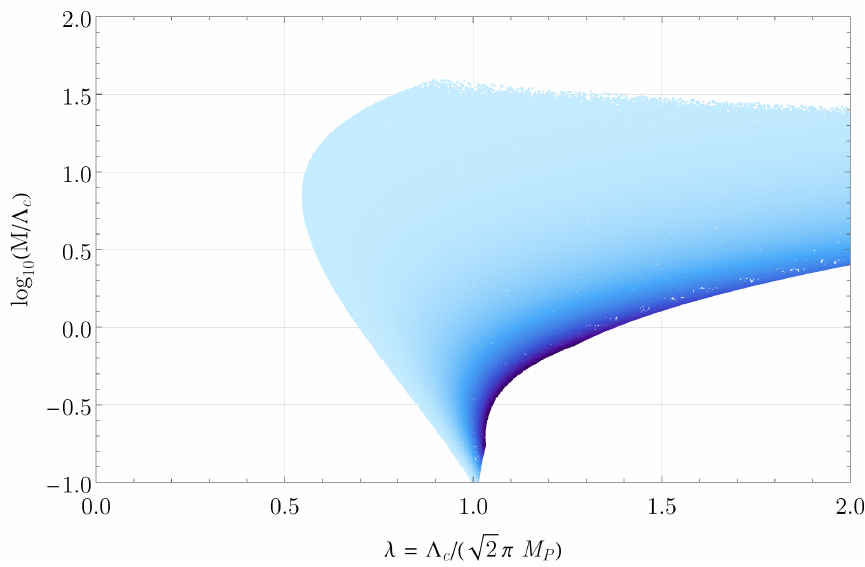}
  \includegraphics[trim={0 0cm 0 0},clip,width=.1\linewidth]{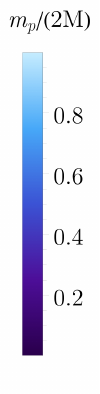}
\caption{
Range of solutions allowed by the model describing pseudoscalar composite particle at chiral limit.}
\label{mass_mp}
\end{figure}

The concrete model realized in the previous sections naturally accommodates this picture. Although the constituent fermion mass $m_Q \sim M$ is large, the confinement scale $\Lambda_D \sim \Lambda_c$ can be parametrically smaller over a wide region of parameter space. As a result, the relic abundance is fixed by heavy degrees of freedom prior to confinement, while late-time interactions are controlled by a much lower dynamical scale. Our numerical results illustrating this behavior are shown in Fig.~\ref{mass_mp} for the pseudoscalar meson sector.

\section{Discussion and Conclusions}\label{sec:conclu}

In this work we have shown that dynamical mass generation can arise as a direct consequence of the gauge-geometric structure underlying models of unconventional supersymmetry. Starting from a single gauge connection valued in the super Lie algebra $su(2,2|3)$, which unifies gravity, $SU(3)$ Yang--Mills fields, and fermionic matter, we derived an effective low-energy theory in which four-fermion interactions of the Nambu--Jona-Lasinio type emerge without introducing elementary scalars or ad hoc couplings.

The key feature of this construction is that all interaction terms, including the quartic fermion vertices, are fully determined by the algebraic structure and gauge invariance of the theory. The NJL coupling constant is fixed to be $G = 1/(6M_P^2)$, a value dictated by the underlying geometry rather than by phenomenological fitting. Likewise, the effective bare fermion mass $m_{\mathrm{eff}}$ is generated by the curvature of the anti--de Sitter background through a gravitational Pauli-like coupling, linking the origin of mass directly to the geometry of spacetime.

A distinctive aspect of the model is that, for $N=3$ colors, the tensor channel in the four-fermion interaction vanishes identically as a consequence of the Fierz identity. This leaves a purely scalar--pseudoscalar NJL interaction, which is precisely the structure needed to drive chiral symmetry breaking and dynamical mass generation. This cancellation is not imposed by hand but follows from the choice of gauge group dictated by the superalgebra.

The analysis of the mass gap equation reveals that, due to the extremely small NJL coupling, the cutoff $\Lambda_c$ must be of the order of the Planck mass in order to generate constituent masses of the order of GeV. In dimensionless terms, this translates into the requirement $\lambda \equiv \Lambda_c/(\sqrt{2}\pi M_P) \simeq 1$, with deviations from unity that must satisfy $|\lambda - 1| \ll 10^{-22}$ to produce phenomenologically relevant masses. While this represents a severe fine-tuning, it is worth noting that the model contains only two free parameters---$\xi$ and $\Lambda_c$---both of which are constrained by the geometric construction. In the limit $\lambda = 1$, the effective bare mass becomes negligibly small ($m_{\mathrm{eff}} \sim 10^{-42}$~MeV), effectively realizing the chiral limit.

The composite particle spectrum was analyzed via the Bethe--Salpeter equation in the random phase approximation. For a constituent mass $M = 340$~MeV, the model predicts a pseudoscalar bound state with mass $m_p \simeq 503$~MeV, while no stable scalar state is found. The absence of a real solution in the scalar channel is consistent with the expectation that the scalar meson lies above the two-particle threshold and appears as a broad resonance rather than a bound state.

The pseudoscalar mass prediction, although obtained within a single-flavor toy model, falls in the range characteristic of hadronic physics. This suggests that the geometric mechanism studied here captures essential qualitative features of chiral symmetry breaking, even though a quantitative comparison with QCD would require extending the model to multiple flavors and incorporating confinement effects beyond the NJL approximation.

We have also explored the implications of this framework for dark matter phenomenology within the gluequark paradigm. In this scenario, the heavy colored fermions arising from the gauge multiplet undergo thermal freeze-out at high temperatures and subsequently bind into neutral composite states upon confinement. The stability of these states may follow from residual symmetries inherent to the algebraic construction, without requiring an ad hoc $Z_2$ parity. The model naturally accommodates the hierarchy $m_Q \gg \Lambda_D$ needed for viable gluequark dark matter, and the pseudoscalar composite states identified in Sec.~\ref{sec:masscomposite} provide concrete candidates within this paradigm.

Several directions for future investigation are worth highlighting. First, the inclusion of additional fermion flavors, which would require enlarging the superalgebra, could allow for a more realistic description of the hadronic spectrum and enable the study of flavor symmetry breaking patterns. Second, the role of torsion deserves further exploration: as noted in Sec.~\ref{sec:model}, the antisymmetric component of torsion couples to the axial fermion bilinear and can be interpreted as an effective chiral chemical potential when a condensate forms. Incorporating a nontrivial torsion background into the gap equation could modify the phase structure of the model and lead to new phenomena, such as parity-violating condensates. Third, the extreme sensitivity of the constituent mass to the cutoff parameter $\lambda$ raises the question of whether a more fundamental mechanism---such as a renormalization group flow or a dynamical determination of the cutoff---could explain the proximity of $\lambda$ to unity.

Finally, it is worth emphasizing the conceptual economy of the approach. In contrast to models that introduce separate mechanisms for symmetry breaking, mass generation, and dark matter stabilization, the present construction derives all of these features from a single gauge-geometric principle. The fact that a Planck-scale NJL coupling, combined with a Planck-scale cutoff, can produce GeV-scale constituent masses and sub-GeV composite states illustrates how large hierarchies can emerge naturally from the interplay between geometric structure and non-perturbative dynamics. Whether this mechanism can be extended to a fully realistic model of particle physics remains an open and compelling question.

\section*{Acknowlegements}

P.A. acknowledges support from ANID-FONDECYT Regular grant No. 1230112.
J.Z.'s work has been partially supported by grant 1180368 from Fondecyt/ANID-Chile.
C.V acknowledge finantial support drom Fondecyt/Anid project 1250206.
\begin{appendices}

\setcounter{equation}{0}
\renewcommand{\theequation}{\thesection\arabic{equation}}
\section{Curvature components}\label{AppA} 

Throughout this section, the fields are the non-canonically normalized ones, as in \eqref{gaugeconnection}. When switching to the canonically normalized Dirac spinors in \eqref{action}, we use a scaling
\begin{eqnarray}
    \psi ^{\mathrm{physical}} = \mu \psi
\end{eqnarray}
where $\mu$ has to have appropriate physical dimensions. We subsequently drop the label ``physical'' from the spinors to simplify the notation.

Here we give the explicity expresions of the curvature components. The field strength is defined by
\begin{equation}
\FF=\half \calF^{ab} \JJ_{ab}+\calF^a \JJ_a +\calG^a \KK_a +\calH \DD +\calF^I \TT_I +\calF \ZZ +\QQb_\alpha^i \calX_i^\alpha +\ocalX_\alpha^i \QQ_i^\alpha\,,
\end{equation}
\begin{align}
\calF^{ab}&=\mathcal{R}^{ab}-\overline{\psi}^i\se\Sigma^{ab}\se\psi_i\,\\
\calF^a&=Df^a+g^a h+\frac{1}{2}\overline{\psi}^i\se\gamma^a\se\psi_i\,,\\
\calG^a&=Dg^a+f^ah-\half \overline{\psi}^i\se\tilde{\gamma}^a\se\psi_i\,,\\
\calH  &=H+f^ag_a+\half \overline{\psi}^i\se\gf \se\psi_i\,,\\
\calF^I&= F^I-i\overline{\psi}^i\se(\lambda^I)_i^{\ j}\se\psi_j\,,\\
\calF &=F-\frac{i}{4 z}\overline{\psi}^i\se\se\psi_i\,,\\
\calX_i^\alpha &=D(\se \psi_i)^\alpha+\frac{1}{2}f^a (\gamma_a\se \psi_i)^\alpha +\half g^a(\tilde{\gamma}_a\se \psi_i)^\alpha+\half h (\gf \se\psi_i)^\alpha\,,\\
\ocalX^i_\alpha&=-(\overline{\psi}^i\se)_\alpha\overleftarrow{D}+\frac{1}{2} (\overline{\psi}^i\se\gamma_a)_\alpha f^a +\half (\overline{\psi}^i\se\tilde{\gamma}_a)_\alpha g^a+\half (\overline{\psi}^i\se\gf )_\alpha h\,,
\end{align}
where
\begin{align}
H&=dh\,,\\
\mathcal{R}^{ab}&=R^{ab}+f^af^b-g^ag^b\,,\\
 R^{ab}&=d\omega^{ab}+\omega^a_{\ c}\omega^{cb}\,,\\
 F^I&=dA^I+\frac{1}{2} f_{JK}^{I}A^J A^K\,,\\
 F&=dA\,,\\
 D_{\text{(Lorentz)-vector}}V^a&=dV^a+\omega^a_{\ b}V^b\\
 D_{\text{spinor}}\psi^\alpha&=d\psi^\alpha+\half \omega^{ab}(\Sigma_{ab}\psi)^\alpha-\frac{i}{2}A^I(\lambda_I\psi)_i^\alpha-i (4/N-1)A\psi_i^\alpha\,,
\end{align}

The covariant derivative $D$ is defined for the $SO(1,3)\times SU(N)\times U(1)$ connection. The left-acting exterior derivative satisfies $\Omega^m\overleftarrow{d}=(-1)^m d\Omega^m$ for an $m$-form, in the spinor representation we get relations (\ref{covD}).

\section{$su(2,2|3)$ algebra}\label{AppB}

Let us consider the following representation of $su(2,2|3)$
\begin{eqnarray}
&\JJ_a =\left[\begin{array}{c|c}
\frac{1}{2}\gamma_a &  0_{4\times3}\\[0.5em] \hline
0_{3\times4} & 0_{3\times3} \\
\end{array}\right]\,, \quad \text{or} \quad (\JJ_a)^A_{\ B}=\frac{1}{2}(\gamma_a)^\alpha_{\ \beta}\delta^A_{\ \alpha} \delta^\beta_{\ B}=\frac{1}{2}(\gamma_a)^A_{\ B}\,,&\\
&\JJ_{ab} =\left[\begin{array}{c|c}
\frac{1}{4}[\gamma_a,\gamma_b] &  0_{4\times3}\\[0.5em] \hline
0_{3\times4} & 0_{3\times3} \\
\end{array}\right]\,, \quad \text{or} \quad (\JJ_{ab})^A_{\ B}=\frac{1}{4}[\gamma_a,\gamma_b]^A_{\ B}=(\Sigma_{ab})^A_{\ B}\,,&\\
&\KK_a =\left[\begin{array}{c|c}
\frac{1}{2}\tilde{\gamma}_a &  0_{4\times3}\\[0.5em] \hline
0_{3\times4} & 0_{3\times3} \\
\end{array}\right]\,, \quad \text{or} \quad (\KK_a)^A_{\ B}=\frac{1}{2}(\tilde{\gamma}_a)^A_{\ B}\,,&\\
&\DD =\left[\begin{array}{c|c}
\frac{1}{2}\gamma_5 &  0_{4\times3}\\[0.5em] \hline
0_{3\times4} & 0_{3\times3} \\
\end{array}\right]\,, \quad \text{or} \quad (\DD)^A_{\ B}=\frac{1}{2}(\gamma_5)^A_{\ B}\,,&\\
&\TT_{I} =\left[\begin{array}{c|c}
0_{4\times4} &  0_{4\times3}\\[0.5em] \hline
0_{3\times4} & \frac{i}{2}\lambda_{I}^{t} \\
\end{array}\right]\,, \quad \text{or} \quad (\TT_I)^A_{\ B}=\frac{i}{2}(\lambda_I^{t})_{\ B}^{A} \,,&\\
&(\QQ^\alpha_i)^A_{\ B}=\left[\begin{array}{c|c}
0_{4\times4} & 0_{4\times 3}\\ [0.5em] \hline
\delta^A_j \delta^\alpha_B & 0_{3\times3}
\end{array}\right]=\delta^A_j \delta^\alpha_B\,,&\\
&(\QQb_\alpha^i)^A_{\ B}=\left[\begin{array}{c|c}
0_{4\times4} & \delta^A_\alpha \delta^i_B\\ [0.5em] \hline
0_{3\times 4} & 0_{3\times3}
\end{array}\right]=\delta^A_\alpha \delta^i_B\,,&\\
&\ZZ^A_{\ B}=\left[\begin{array}{c|c}
i\delta^\alpha_\beta &0_{4\times3}\\ [0.5em] \hline
0_{3\times4} & \frac{4}{3} i\delta^i_j\end{array}\right]=
\left(i\delta^A_\beta\delta^\beta_B+\frac{4i}{3}\delta^A_j\delta^j_B\right)\,,&
\end{eqnarray}
where $\gamma_5=i\gamma^0 \gamma^1 \gamma^2 \gamma^3$, $(\gamma_5)^2=\mathds{1}$ and 
\begin{align}
\tilde{\gamma}_a &=\frac{i}{3!}\epsilon_{abcd}\gamma^{bcd}=-\gamma_5\gamma_a\,,\\
\gamma_{abc}    &=  \gamma_{[a}\gamma_b \gamma_{c]}=\frac{1}{3!}\sum_{\Pi(a,b,c)} \text{sign}(\Pi(a,b,c)) \gamma_a \gamma_b \gamma_c=i\epsilon_{abcd}\gamma_5 \gamma^{d}\,.
\end{align}

The $\gamma$-matrices are in a $4\times 4$ spinor-representation ($\alpha, \beta,\cdots$ run from 1 to 4). The indexes of the tangent space $a,b=0,1,2,3$. Indexes in the adjoint representation of $su(3)$ take values $I,J=1,2,\ldots,8$, and in the fundamental take the values $i,j=1,2$. The $\gamma$-matrices are endomorphisms and they act on spinors
\begin{equation}
 \psi^\alpha \stackrel{\gamma_a}{\longrightarrow} (\gamma_a)^\alpha_{\ \beta} \psi^\beta\,.
\end{equation}
These $\gamma$-matrices satisfy $\{\gamma^a,\gamma^b\}=2 \eta^{ab}$, where the metric $\eta$ is given by $\eta=\mathrm{diag}(-,+,+,+)$. The spinor indexes will be often omitted.

In a similar way the $\lambda$-matrices are also endomorphisms and they act on spinors as
\begin{equation}
 \psi^\alpha_i \stackrel{\lambda_I}{\longrightarrow} (\lambda_I)_i^{\ j}\psi^\alpha_j \,.
\end{equation}
The $\lambda$-matrices satisfy $[\lambda_I,\lambda_J]=f^{IJK} \lambda_K$, where indexes are raised/lowered with an Euclidean metric $\delta_{IJ}$. Indexes of the representation are $A,B=1,\cdots,7$, so we have a $7\times7$ representation. We find convenient to split $A=(\alpha,i)$. All the possible products that mix spaces like $p^i_{\ A}  q^A_{\ \alpha}$ are trivial. Thus, the following relations are understood
\begin{eqnarray}
&(\gamma_a)^A_{\ B}=\delta^A_\alpha (\gamma_a)^\alpha_{\ \beta} \delta^\beta_B\,,&\\
&C_{\alpha A}=C_{\alpha\beta} \delta^\beta_A \,,&
\end{eqnarray}

The generators $\JJ_a$ and $\JJ_{ab}$ form a adS$_4$ algebra,
\begin{align}
& [\JJ_a,\JJ_b]=\JJ_{ab}\,,\\
& [\JJ_a,\JJ_{bc}]=\eta_{ab}\JJ_c-\eta_{ac}\JJ_b\,,\\
& [\JJ_{ab},\JJ_{cd}]=-(\eta_{ac}\JJ_{bd}-\eta_{ad}\JJ_{bc}-\eta_{bc}\JJ_{ad}+\eta_{bd}\JJ_{ac})\,.
\end{align}
Along $\DD$ and $\KK_a$ they form the conformal algebra,
\begin{align}
& [\KK_a,\KK_b]=-\JJ_{ab}\,,\\
& [\JJ_a,\KK_{b}]=\eta_{ab}\DD\,,\\
& [\KK_a,\JJ_{bc}]=\eta_{ab}\KK_c-\eta_{ac}\KK_b\,,\\
& [\DD,\KK_{a}]=-\JJ_a\,,\\
& [\DD,\JJ_{a}]=-\KK_a\,.
\end{align}

For the internal generators we have the $su(3)$ algebra
\begin{equation}
 [\TT_I,\TT_J]=f^{IJK}\TT_K\,,
\end{equation}
and they are anti-hermitian $\TT_I^\dag=-\TT_I$ (also $\ZZ^\dag=-\ZZ$).

Inlcuding $\QQ^\alpha_i$ and $\QQb^i_\alpha$ the commutators close in a $su(2,2|3)$ superalgebra
\begin{eqnarray}
&[\JJ_a,\QQb_\alpha^i]=\frac{s}{2}\QQb_\beta^i(\gamma_a)^\beta_{\ \alpha}\,, \quad [\JJ_a,\QQ^\alpha_i]=-\frac{s}{2}(\gamma_a)^\alpha_{\ \beta}\QQ^\beta_i\,,&\\
&[\JJ_{ab},\QQb_\alpha^i]=\QQb_\beta^i(\Sigma_{ab})^\beta_{\ \alpha}\,, \quad [\JJ_{ab},\QQ^\alpha_i]=-(\Sigma_{ab})^\alpha_{\ \beta}\QQ^\beta_i\,,&\\
&[\KK_a,\QQb_\alpha^i]=\frac{1}{2}\QQb_\beta^i(\tilde{\gamma}_a)^\beta_{\ \alpha}\,, \quad [\KK_a,\QQ^\alpha_i]=-\frac{1}{2}(\tilde{\gamma}_a)^\alpha_{\ \beta}\QQ^\beta_i\,,&\\
&[\DD,\QQb_\alpha^i]=\frac{1}{2}\QQb_\beta^i(\gamma_5)^\beta_{\ \alpha}\,, \quad [\DD,\QQ^\alpha_i]=-\frac{1}{2}(\gamma_5)^\alpha_{\ \beta}\QQ^\beta_i\,,&\\
&[\TT_I,\QQb_\alpha^i]=-\frac{i}{2}\QQb_\alpha^j (\lambda_I)^{\ i}_j\,, \quad [\TT_I,\QQ^\alpha_i]=\frac{i}{2}(\lambda_I)^{\ j}_i\QQ^\alpha_j\,,&\\
&[\ZZ,\QQb_\alpha^i]=-\frac{i}{3}\QQb_\alpha^i\,, \quad [\ZZ,\QQ^\alpha_i]=\frac{i}{3}\QQ^\alpha_i\,,&
\end{eqnarray}
\begin{equation}
\begin{split}
\{\QQ^\alpha_i,\QQb_\beta^j\}=\left(\frac{1}{2}(\gamma^a)^\alpha_{\ \beta} \JJ_a-\frac{1}{2}(\Sigma^{ab})^\alpha_{\ \beta} \JJ_{ab}-\frac{1}{2}(\tilde{\gamma}^a)^\alpha_{\ \beta} \KK_a+\frac{1}{2}(\gamma^5)^\alpha_{\ \beta} \DD\right)\delta^j_i
\\
+\delta^\alpha_\beta\left(-i(\lambda_I)_i^{\ j}\TT_I-\frac{i}{4}\delta_i^{j} \ZZ\right)\,.
\end{split}
\end{equation}


\end{appendices}

\bibliographystyle{ieeetr}
\bibliography{main.bib}

\end{document}